\documentclass[sigconf,table]{acmart}

\renewcommand\footnotetextcopyrightpermission[1]{} %
\usepackage{color, colortbl}
\usepackage{url}
\usepackage{multirow}
\usepackage{hyperref}

\usepackage[pscoord]{eso-pic}%

\acmPrice{15.00}
\acmDOI{10.1145/123456.123456}  %
\acmISBN{978-1-4503-9999-9/19/11} %
\copyrightyear{2020}
\acmYear{2020}
\setcopyright{acmlicensed}
\acmConference[]{}
\acmBooktitle{}

\newcommand{\preprintBanner}{
\AddToShipoutPictureFG*{\put(\LenToUnit{0.5\paperwidth},\LenToUnit{0.95\paperheight}){\makebox[0pt][c]{
\renewcommand{\arraystretch}{1.5}
\setlength{\tabcolsep}{18pt}
\rowcolors{1}{Preprint}{Gray}
\begin{tabular}{|p{0.55\paperwidth}|} \hline
\textbf{Preprint from \texttt{\href{https://ostendorff.org/pub/}{https://ostendorff.org/pub/}}} \\ \hline
\footnotesize
M. Ostendorff, T. Ruas, M. Schubotz, G. Rehm, B. Gipp, ``Pairwise Multi-Class Document 
Classification for Semantic Relations between Wikipedia Articles'' in \textit{Proceedings of the ACM/IEEE Joint Conference on Digital Libraries (JCDL)}, 2020. \\ \hline
\end{tabular}}}}
}

\newcommand{\relCount}{nine~}

\definecolor{Gray}{gray}{0.925} %
\definecolor{Preprint}{rgb}{.63,.79,.95}

\begin{document}

\title{Pairwise Multi-Class Document Classification for Semantic Relations between Wikipedia Articles}

\author{Malte Ostendorff${}^{1,2}$,Terry Ruas${}^{3}$, Moritz Schubotz${}^{3}$, Georg Rehm${}^{1}$, Bela Gipp${}^{2,3}$}

\affiliation{
	\institution{
    \textsuperscript{1}DFKI GmbH, Germany (firstname.lastname@dfki.de)
    }
}

\affiliation{
	\institution{
    \textsuperscript{2}University of Konstanz, Germany (firstname.lastname@uni-konstanz.de)
    }
}

\affiliation{
	\institution{
    \textsuperscript{3}University of Wuppertal, Germany (lastname@uni-wuppertal.de)
    }
}

\renewcommand{\shortauthors}{Ostendorff et al.}

\begin{abstract}

Many digital libraries recommend literature to their users considering the similarity between a query document and their repository. However, they often fail to distinguish what is the relationship that makes two documents alike. In this paper, we model the problem of finding the relationship between two documents as a pairwise document classification task. 
To find the semantic relation between documents, we apply a series of techniques, such as GloVe, Paragraph-Vectors, BERT, and XLNet under different configurations (e.g., sequence length, vector concatenation scheme), including a Siamese architecture for the Transformer-based systems.
We perform our experiments on a newly proposed dataset of 32,168 Wikipedia article pairs and Wikidata properties that define the semantic document relations.
Our results show vanilla BERT as the best performing system with an F1-score of 0.93,
which we manually examine to better understand its applicability to other domains.
Our findings suggest that classifying semantic relations between documents is a solvable task and motivates the development of recommender systems based on the evaluated techniques.
The discussions in this paper serve as first steps in the exploration of documents through SPARQL-like queries such that one could find documents that are similar in one aspect but dissimilar in another.

\end{abstract}



\maketitle

\preprintBanner

\section{Introduction}

To cope with the ever-emerging information overload, digital libraries employ literature recommender systems (LRS)~\cite{Beel2016}.
These systems recommend related documents with the help of similarity measures, which often only distinguish between similar and dissimilar documents. 
This simplification neglects the many facets of extensive documents in digital libraries.
It remains unclear to which of the many facets the similarity relates.
In philosophy~\cite{Goodman1972}, but also in natural language processing (NLP)~\cite{Bar2011}, the similarity of A to B has been addressed as an ill-defined notion unless one can say to what the similarity relates.
For LRS, one would rather know what aspects of the two documents are similar or how they relate to each other than just knowing that the documents are similar or dissimilar.
Identifying the aspects connecting different documents would allow users to explore the document space by formulating SPARQL-like queries in terms of documents and their relations (e.g., find a document with one specific relation to A, but a different relation to B).
These queries are generally referred to as analogical queries~\cite{Gick1983}.
Especially for complex information needs, the formulation of analogical queries is more intuitive~\cite{Lofi2017}.
A system that supports analogical queries would be particularly beneficial for scientific literature since the discovery of the analogies is crucial for scientific progress~\cite{Chan2018}.

Nonetheless, document similarity measures do not take into account the semantic relations that would underpin such a system.  
While other NLP tasks, like relation extraction (RE)~\cite{Yao2019}, deal with relations, they are not concerned with semantic relations between documents.
For instance, RE is about relations between entities occurring within a single document text.
Similarly, the document classification task aims to categorize individual documents, but fail to address the relationship that binds two or more documents.

\begin{figure}[h]
\centering
\includegraphics[page=11,clip,width=0.49\textwidth,trim={2.4cm 3cm 2cm 2cm},clip]{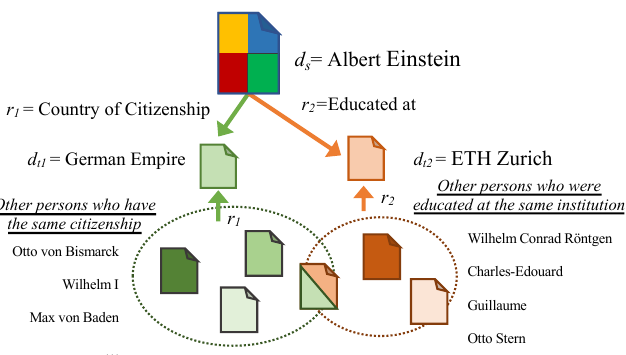}
\caption{\label{fig:wikidata-usecase}Semantic relations between Wikipedia articles. Seed article \textit{Albert Einstein} is connected to other articles by the property \textit{educated at} and \textit{citizenship}. Considering articles only a single edge apart leads to diverse recommendations, while two edges can be utilized for recommendation sets focused on a specific or an intersection of aspects.}
\end{figure}

In this paper, we combine the ideas of relation extraction, document classification, and document similarity to classify the semantic relation of document pairs.  
Given a seed document $d_s$, we are interested in finding a target document $d_t$ that shares the semantic relation $r_i$ with $d_s$.
We use the term ``semantic relation'' to indicate connections between two documents above the syntax level~\cite{Khoo2007}.
We model the task of finding the relation $r$ of a document pair $(d_s,d_t)$ as a pairwise multi-class document classification problem. 
The semantic relation between documents provides context for similarity and enables analogical queries.
To evaluate the presented techniques, we build a dataset using Wikipedia and Wikidata~\cite{Vrandecic2014} repositories to illustrate our problem. 
Wikipedia articles are the seed and target documents, while Wikidata properties provide the semantic relations between a document pair. 
Figure~\ref{fig:wikidata-usecase} shows one example from our dataset.
The articles \textit{Albert Einstein} and \textit{German Empire} are the pair $(d_s,d_{t_1})$ and the relation is defined by $r_1$, which is the Wikidata property \textit{country of citizenship}.
These relations enable recommendations and analogical queries (Section~\ref{sssec:rel2rec}).

Our paper makes three major contributions. 
First, we propose a method to classify the semantic relations of document pairs. 
Second, we implement six different models using word-based document embeddings from GloVe~\cite{Pennington2014} and Paragraph Vectors~\cite{Le2014} (as Doc2vec implementation~\cite{gensim}), and deep contextual language models from BERT~\cite{Devlin2019} and XLNet~\cite{Yang2019} in a vanilla and Siamese architecture~\cite{Bromley1993}. Each system is evaluated under specific configurations regarding its concatenation method and sequence length. 
Third, we introduce a novel dataset composed of 32,168 Wikipedia article pairs and Wikidata properties that define the semantic relation of these articles. 
All our datasets, trained models, and source code are publicly available to contribute to transparency and reproducibility. %

\section{Related Work}
\label{sec:related-work}

In the following, we relate to other research regarding document similarity and its use for recommender systems and analogies. 
Besides, we refer to related work that applies similar or the same techniques for solving other NLP tasks.

\subsection{Document Similarity \& Recommendations}

B\"{a}r et al.~\cite{Bar2011} discuss the notion of similarity between texts in the context of NLP. 
They express that while text similarity is present in many NLP tasks, the similarity is often ill-defined and used as an ``umbrella term covering quite different phenomena''. 
B\"{a}r et al.~\cite{Bar2011} formalize text similarity and suggest content, structure, and style as the major dimensions inherent to texts. 
With approximately 55\% of publications using content-based filtering, it accounts for the majority of the LRS research~\cite{Beel2016}. 
Structure and style are not actively being accounted for. 
Therefore, we focus only on the content.

Giving its diversity and reach, Wikipedia had been used as a laboratory in which recommender system methodologies can be tested~\cite{Ollivier2007, Schwarzer2016}. 
In ~\cite{Schwarzer2016}, we compared text- and link-based document similarity measures and found that both methods capture similarity differently.
Link-based methods tend to retrieve documents from a broader context, while text-based methods are focused on specific terms and topics. Consequently, each similarity approach is suitable for different information needs, e.g., getting an overview of a topic or performing in-depth research. With the classification of semantic document relations, we intend to tailor recommendations depending on specific information needs. 
For example, we could provide either recommendations focusing on a particular relation class or diverse recommendations from multiple relation classes.

\subsection{Analogical Queries}

An analogy is a comparison between two or more elements in which their relation is used to illustrate an explanation.
Moreover, analogical query solving in the form of ``A is to B as C is to ?'' is a fundamental aspect of human intelligence~\cite{Gick1983,Lofi2017}.  %
Chan et al.~\cite{Chan2018} emphasize the importance of analogical query solving for scientific progress. 
They propose a semi-automated approach for finding analogies between research papers using expert and crowd annotators to segment the abstracts of papers into background, purpose, mechanism, and findings. 
Next, they encode the segments with GloVe~\cite{Pennington2014} and Paragraph Vectors~\cite{Le2014} and compute their similarity to determine whether papers are similar with respect to those segments. However, segmentation breaks the coherence of documents. Our method aims to find semantic relations between documents while maintaining their coherence intact.
In the context of word embeddings, analogies are often illustrated using vector arithmetic, e.g., $\vec{w}_{\text{King}}-\vec{w}_{\text{Queen}}=\vec{w}_{\text{Man}}-\vec{w}_{\text{Woman}}$~\cite{Mikolov2013}. Allen and Hospedales~\cite{Allen2019} give a mathematical description of analogies as linear relationships between word embeddings. 
Dai et al.~\cite{Dai2015} demonstrate that such analogies are also present in document embeddings. 
In their experiment, using Wikipedia articles, the nearest neighbor to the vector of $\vec{w}_{\text{LadyGaga}}-\vec{w}_{\text{American}}+\vec{w}_{\text{Japanese}}$ is the article on Ayumi Hamasaki, a famous Japanese singer that published an album called ``Poker Face'' in 1998 (like Lady Gaga in 2008).

\subsection{Transformers}
Recently, Transformer-based~\cite{Vaswani2017} neural language models introduced a shift from context-free word embeddings, like GloVe~\cite{Pennington2014}, to contextual embeddings as the ones used in BERT~\cite{Devlin2019} and XLNet~\cite{Yang2019}.
The Transformer architecture allowed the efficient unsupervised pretraining of language models and led to significant improvements in many NLP benchmarks~\cite{Zhu2015,Wang2019,Misra2016}.
Reimers and Gurevych~\cite{Reimers2019} proposed to combine BERT with a Siamese architecture~\cite{Bromley1993} for semantic representations of sentences and their similarity~\cite{Misra2016}.
In prior work~\cite{Rehm2020c}, we also utilized a Siamese BERT model to determine the discourse relations between text segments to generate a story for the segments.
Moreover, BERT has successfully solved various document classification tasks~\cite{Adhikari2019,Ostendorff2019}. %
Akkalyoncu Yilmaz et al.~\cite{AkkalyoncuYilmaz2019} apply BERT to an information retrieval system for an end-to-end search over large document collections.
Despite their success in NLP, Transformers have gained little attention in the recommender system community so far and are not even mentioned in a recently published survey~\cite{Bai2019}. 
To our knowledge, Hassan et al.~\cite{Hassan2019} are one of the first to use BERT to recommend research papers. As opposed to our work, Hassan et al. use BERT to encode only the paper titles as vectors and then generate recommendations using cosine similarity. In our experiments, we utilize the article text and learn the document relation using a multilayer perceptron (MLP).

\section{Methodology}

In the following, we describe the dataset and investigated systems to facilitate the reproduction of our results.

\subsection{Data set \& Use case} \label{ssec:dataset}

Existing datasets provide either classifications of single documents (e.g., topic~\cite{Ostendorff2019}), relations between sentences or entities (e.g., natural language inference~\cite{Wang2019}, word analogies~\cite{Mikolov2013}, entity relation extraction~\cite{Yao2019}), or similarity between text pairs (i.e., binary classification~\cite{Dolan2005}). 
Our task is defined as as multi-class classification of document pairs consisting of multiple sentences. 
Moreover, the learning characteristic in our task requires considerably larger dataset than~\cite{Chan2018} or~\cite{Jaradeh2019}. 
To the best of our knowledge, no established dataset fulfills these requirements.

\subsubsection{Training data} \label{sssec:trainingdata}

One example of a digital library that employs an LRS is Wikipedia. 
Recommendations for Wikipedia articles have been addressed in the literature~\cite{Schwarzer2016,Ollivier2007}. Wikipedia is connected with Wikidata, an open knowledge graph in which nodes represent items (e.g., Wikipedia articles) and edges represent properties of these items (e.g., relation that connect two different articles). 
The link of most Wikipedia articles to their corresponding Wikidata items allows the construction of a large dataset tailored to the problem of semantic relation classification. 
The triple $(d_s,d_t,r_i)$ of two documents $d_s$ and $d_t$, and the relation class $r_i$ describes a document pair relation. 
In the Resource Description Framework (RDF) terminology, $d_s$ is the subject, $d_t$ the object, and $r_i$ the predicate, whereas in the Wikidata terminology, a relation corresponds to a statement\footnote{\url{https://www.wikidata.org/wiki/Help:Statements}}. 
The relation class $r_i$ (predicate) is a Wikidata property that semantically relates a pair of Wikipedia articles $(d_s,d_t)$.
For instance, the Wikipedia article of \textit{Albert Einstein}\footnote{\url{https://en.wikipedia.org/wiki/Albert\_Einstein}} and its Wikidata item\footnote{\url{https://www.wikidata.org/wiki/Q937}} is connected to the article\footnote{\url{https://en.wikipedia.org/wiki/German\_Empire}} and item\footnote{\url{https://www.wikidata.org/wiki/Q43287}} of the \textit{German Empire} through the property \textit{country of citizenship}\footnote{\url{https://www.wikidata.org/wiki/Property:P27}}.
The Wikidata property acts as both, the relation of the Wikipedia article pair and the class label in the training data for this same pair of documents. Table~\ref{tab:relations} lists other examples to illustrate our scenario better.

Given Wikipedia's nature as an encyclopedia, its use as the dataset has some shortcomings. 
Encyclopedic documents tend to describe a single entity, and their semantics can be seen as rather homogeneous in comparison to other literature forms. Nonetheless, we consider Wikipedia and Wikidata to be a suitable corpus to demonstrate our approach. Wikidata properties range from entity-specific relations (e.g., \textit{educated at}) to abstract ones (e.g., \textit{facet of}). Wikipedia articles and their relations are, on average, more comprehensible than those in scientific literature
, which contributes to the analysis of our results. 
Another aspect that supports our choice of Wikipedia and Wikidata is their open license copyright.

\subsection{Semantic Relations}
\label{ssect:semantic-relations}

At the time of writing, Wikidata contained 7,091 properties\footnote{\url{https://tools.wmflabs.org/hay/propbrowse/}} of which we selected the following \relCount for this research:

\begin{itemize}
  
    \item \textit{country of citizenship} - seed is citizen of the target;
      \item \textit{different from} - item that is different from another item, with which it is often confused;
      \item \textit{educated at} - educational institution attended by seed;
      \item \textit{employer} - seed works or worked for target; %
    
    \item \textit{facet of} - topic of which this item is an aspect, item that offers a broader perspective on the same topic;
    
    \item \textit{has effect} - the seed causes the target;
    \item \textit{has quality} - the entity has an inherent or distinguishing non-material characteristic;
    \item \textit{opposite of} - item that is the opposite of this item;
    \item \textit{symptoms} - possible symptoms of a medical condition.
\end{itemize}{}

Table~\ref{tab:relations} lists the corresponding Wikidata PIDs, their quantity, and examples for each property.
Besides the number of available Wikipedia article pairs, diversity was also a criterion in our selection. 
Diversity refers to the different semantic meanings of properties (e.g., \textit{country of citizenship}, \textit{opposite of}). Similarly, the requirements to predict a relation between documents can also be diverse.
While some relations are clearly expressed within the document text (e.g., for documents referencing people, their citizenship is often put in the first sentences), others will require a more comprehensive understanding of the article content. 
For instance, while \textit{floor} as the opposite of \textit{ceiling} is evident, this fact will most likely not be explicitly mentioned in the article text. 
Also, other relations like \textit{has effect} or \textit{symptoms} can require unwritten domain knowledge. 
The classification performance can also be affected by the type of the connected articles.
For example, the relation class \textit{country of citizenship} exclusively connects persons and countries. 
No other property uses such a combination. 
On the contrary, the relation classes \textit{educated at} and \textit{employer}, connect a person with an organization. 
Additionally, all relations are unidirectional, except for \textit{opposite of}. 
Given the many aspects our relations are exploring, we expect significant differences in the classification performance.

\begin{table}[h]
\footnotesize

\begingroup
\setlength{\tabcolsep}{3pt} %
\renewcommand{\arraystretch}{1.6} %
\caption{\label{tab:relations}The relation classes with their Wikidata PIDs, three examples, and the number of samples in our dataset. }
\begin{tabular}{llrp{4.4cm}}
          \textbf{Relation class}
          &    \textbf{PID }
          &  \textbf{\#} 
          & \textbf{Example relations} \\
\midrule
 country of citizenship 
 &    P27 
 &       3636 
 &   Torben Ulrich  $\rightarrow$ Denmak
  \newline Neal Doughty $\rightarrow$ United States
   \newline Julian Kenny $\rightarrow$ Trinidad and Tobago      \\  \hline
 
 different from 
         &  P1889 
         &       4048 
         &   Computer file $\rightarrow$ File folder
         	 \newline  Lee County, Alabama	$\rightarrow$ Lee County, Illinois  
         	 \newline Karo $\rightarrow$ Karo (name)  \\ \hline
         	
            educated at 
            &    P69 
            &       1798 
            &    Hillar Eller  $\rightarrow$ University of Tartu
            \newline Al Young $\rightarrow$	University of Michigan    
            \newline Heinrich Finkelstein $\rightarrow$ 	Leipzig University                            \\  \hline
            
               employer 
               &   P108 
               &       1557 
               &    Gary M. Mavko $\rightarrow$ Stanford University
               \newline Alexander Medvedev $\rightarrow$ Gazprom
               \newline John Reif $\rightarrow$	Duke University     \\  \hline
               
               facet of 
               &  P1269 
               &       1343 
               &   Reformation $\rightarrow$ Protestantism 
               \newline 1974 in Portugal $\rightarrow$ 	Portugal
               \newline  Sportsmanship	$\rightarrow$ Sport    \\  \hline
             has effect 
             &  P1542 
             &        698 
             &   Language attrition $\rightarrow$ Extinct language
             \newline Arsenic poisoning $\rightarrow$	Lung cancer
             \newline Foul ball $\rightarrow$	Out (baseball)     \\  \hline
            has quality 
            &  P1552 
            &       1022 
            &   Antisemitism $\rightarrow$ Nazism    
            \newline Employment	$\rightarrow$ Access badge
            \newline  Human	 $\rightarrow$ Gender                                        \\  \hline
            opposite of &   P461 
            &        929 
            &       Floor $\rightarrow$ Ceiling    
            \newline Person	$\rightarrow$ Society
            \newline  Exponentiation $\rightarrow$ 	Logarithm          \\  \hline
               symptoms &   P780 
               &       1053 
               &    Myalgia $\rightarrow$ Influenza   
               \newline Mercury poisoning	$\rightarrow$ Cough
               \newline  Death rattle $\rightarrow$	Sound                                          \\ \hline

{} & Total & 16,084 & \\

\end{tabular}
\endgroup

\end{table}

\subsection{Data Preprocessing} \label{ssec:data-preprocessing}

We sampled 10,000 article pairs in total with a balanced class distribution over the \relCount properties. 
The relations were obtained trough the Wikidata SPARQL interface in December 2019. 
For each Wikipedia article in the sample, we also checked whether the article was connected to any other article but was not part of the initial sample and retrieved the missing relations. We removed all duplicated article pairs and multi-label relations. The main goal of this paper is to explore the multi-class classification problem, so we ensure that the same pair of documents did not share different labels. Wikidata provides data for multi-label relations, especially for hierarchical properties. However, only less than 1\% of our sample data contained multi-label relations. For the sake of simplicity, we decided to remove them.
This procedure generates 16,084 Wikipedia article pairs with an imbalanced class distribution (Table~\ref{tab:relations}). 
The increase of samples is due to the retrieval of missing relations. 
The corresponding articles were converted to plain-text from the English Wikipedia dump of November 2019 using the Gensim API~\cite{gensim}.
\subsection{Negative Sampling}  \label{ssec:negative-samlping}

In addition to the \relCount positive classes from Wikidata, we introduce a class named \textit{None} that works as negative cases for our positive samples in the same proportion. The articles in the \textit{None} category are randomly selected and do not share any relation with the positive ones. The resulting dataset contains 32,168 samples in total.

\subsection{Systems}
\label{ssec:systems}

This paper evaluates six classifiers under different configurations, totaling 30 systems. 
We distinguish between three model categories: (i) document embeddings from word embeddings using the full document text (GloVe and Doc2vec), (ii) Vanilla Transformers, and  (iii) Siamese Transformers (each Transformer as BERT and XLNet). 
Each classifier takes two documents $d_s$ and $d_t$ as input and predicts their relation $\hat{y}=\text{rel}(d_s,d_t)$ as its output. 
The hyperparameters for the considered systems are detailed in Section~\ref{ssec:hyperparameters}.

\subsubsection{Doc2vec} \label{sssec:doc2vec} %

With word2vec, Mikolov et al.~\cite{Mikolov2013} introduced an algorithm to learn dense vector representations of words such that semantically similar words end up close to each other in the embedding space. 
Word2vec is widely applied in NLP tasks~\cite{iacobacci2016,ruas2019} but unable to represent entire documents.
Paragraph Vectors~\cite{Le2014} (also known as Doc2vec), extends word2vec to learn embeddings for word sequences of arbitrary length. In the following, we refer Paragraph Vectors as Doc2vec, since we employ the widely-used implementation of the Gensim~\cite{gensim} framework. 
We obtained a 200D document vectors $\vec{d}$ for each Wikipedia article by training Doc2vec's distributed bag of words model (dbow) using both training and test data, and the default hyperparameters in Gensim\footnote{\url{https://radimrehurek.com/gensim/models/doc2vec.html}}. 
The document vector size of 200 corresponds to the size of the GloVe word vectors (Section~\ref{sssec:avgglove}).
The choice of dbow over the distributed memory training model is due to its results in semantic similarity tasks~\cite{Lau:16}. 
It is important to mention that even though the embeddings model used both training and test sets, the latter was not used for training the classifier. 

\subsubsection{AvgGloVe} \label{sssec:avgglove}

GloVe~\cite{Pennington2014} also produce dense embedding representations, but unlike word2vec, GloVe is a count-based method that uses global statistics to derive its word vectors. 
In GloVe, the co-occurrence matrix explores the ratio of the probabilities of words in a text to derive its semantic vectors. While we use the 200D pretrained word embedding model\footnote{\url{https://nlp.stanford.edu/projects/glove/}}, GloVe does not provide document vectors directly. 
To embed a Wikipedia article $\vec{d}$, we compute the weighted average over its word vectors $\vec{w_i}$ (AvgGloVe), whereby the number of occurrences of the word $i$ in $d$ defines the weight $c_i$.
Arora et al.~\cite{Arora2017} showed the weighted average of word vectors is effective and yields good results for representing documents. %

For both full-text methods, AvgGloVe and Doc2vec, we encode each document from our document pair $(d_s,d_t)$ independent from the classification task and concatenate their resulting vectors. 
The different concatenation variants tested in our experiments are discussed in Section~\ref{ssec:hyperparameters}. 
The resulting document pair vector is then used as an input to a fully-connected MLP, which classifies the document pair relation $\hat{y}=\text{rel}(d_s,d_t)$. 
The dimension of the output of the last layer of all classifiers ($\hat{y}$), corresponds to the \relCount Wikidata properties (Table~\ref{tab:relations}) and one additional dimension for the \textit{None} class of negative samples (Section \ref{ssec:negative-samlping}). The logistic sigmoid function is used to generate the probabilities for the multi-class classification.

\subsubsection{Vanilla Transformer} \label{ssec:vanilla-transformer}

\begin{figure}[h]
\centering
\includegraphics[page=1,clip,width=0.49\textwidth,trim={2.2cm 2.9cm 2.0cm 4.5cm }]{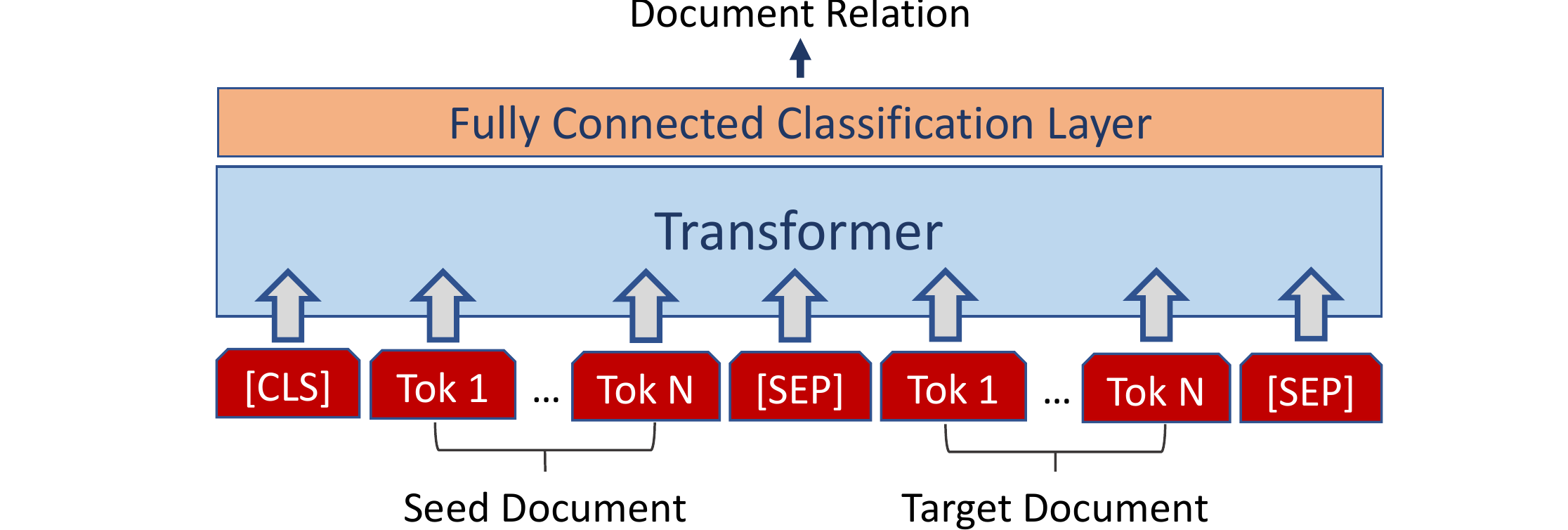}
\caption{\label{fig:transformer}Vanilla Transformer for sequence pair classification. \texttt{[SEP]}-token separates seed and target document.}
\end{figure}

As the third model category, we employ two language models for deep contextual text representations based on the Transformer architecture~\cite{Vaswani2017}, named BERT~\cite{Devlin2019} and XLNet~\cite{Yang2019}. 
The two Transformer models are originally designed to solve sequence pair classification. The base training task (i.e., next sentence prediction) for BERT and XLNet allows us to fine-tune them for the document pair classification task. The content of the document pair (i.e., title and text of $d_s$ and $d_t$) is tokenized, delimited with special tokens, i.e., \texttt{[CLS]} and \texttt{[SEP]} for BERT, \texttt{<cls>} and \texttt{<sep>} for XLNet, and then jointly fed trough the Transformer (Figure ~\ref{fig:transformer}). The Transformer output is used as the input to a single fully-connected linear layer with 512 units for the classification (prediction head). Regarding terminology, we refer to the two models as vanilla Transformer since their original architecture is unchanged.

\subsubsection{Siamese Transformer} \label{sssec:siamese-transformer}

\begin{figure}[h]
\centering
\includegraphics[page=2,clip,width=0.49\textwidth,trim={2cm 2cm 2cm 2cm}]{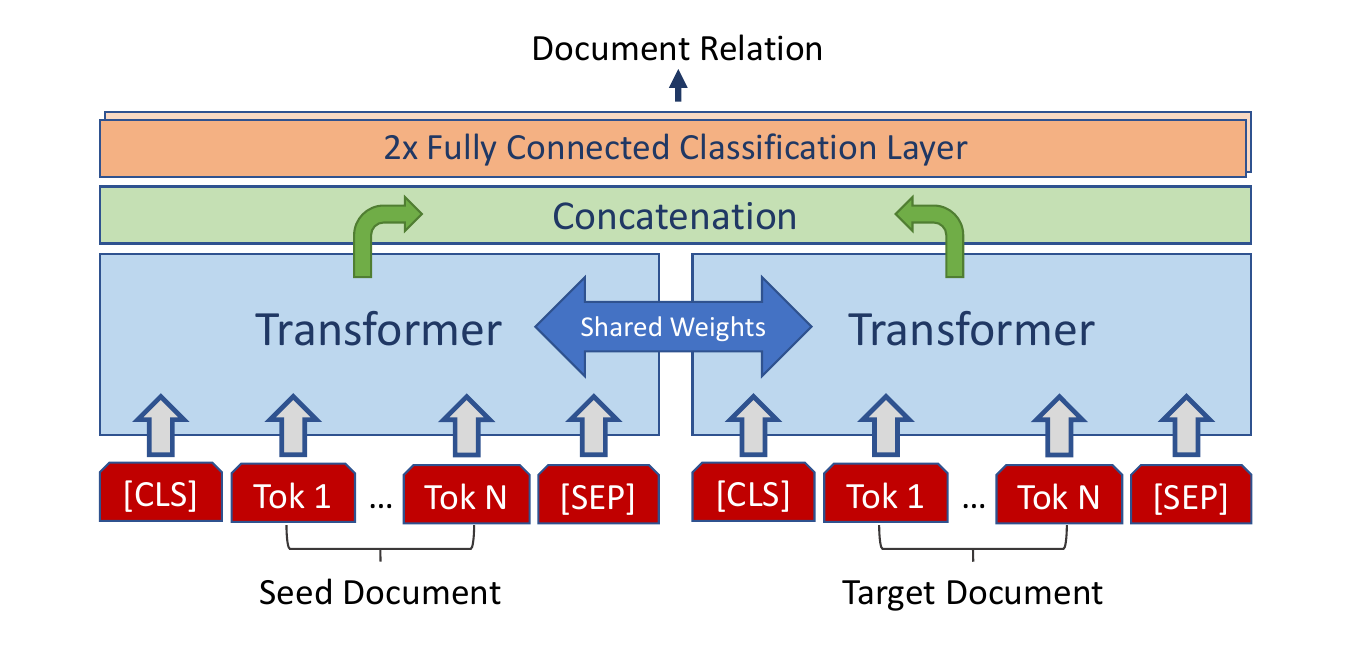}
\caption{\label{fig:siamese}Siamese Transformer architecture. Both documents fed separately trough the Transformer, the concatenated document vectors are input to the classification layer. }
\end{figure}

We combine the two Transformers (BERT and XLNet) in a Siamese network architecture~\cite{Bromley1993}. 
In Siamese networks, two inputs are fed through identical sub-networks with shared weights (in this case, the Transformers), and then passed to a classifier or a similarity function. 
Reimers and Gurevych~\cite{Reimers2019} have shown that Siamese BERT networks are suitable for text similarity tasks. 
For our experiment, both documents $d_s$ and $d_t$ are input to the Transformer sub-networks to derive two contextual document vectors (Figure~\ref{fig:siamese}). 
Next, the document vectors are concatenated and classified with a 2-layer MLP (2x512 units with ReLU activation), the same method applied by Doc2vec and AvgGlove. 
In contrast to Doc2vec and AvgGloVe, the document representations are neither fixed nor frozen, but continually learned during the training of the classifier. 
Different than \cite{Reimers2019}, our implemented Siamese architecture is applied to a multi-class classification instead of a binary one.

The architectures of the underlying BERT and XLNet models are the corresponding \texttt{BASE-CASED} versions of the pretrained models with 12-layer, 768-hidden, 12-head, and 110M parameters. Even though the architectures of BERT and XLNet are comparable, the associated language models are pretrained with different data. While BERT is trained on English Wikipedia and the BooksCorpus~\cite{Zhu2015} alone, XLNet uses additional Web corpora for pretraining \cite{Yang2019}. 

\subsection{Hyperparameters}
\label{ssec:hyperparameters}

\subsubsection{Sequence length} \label{sssec:seq-len}

The vanilla and Siamese Transformer models based on BERT have a maximum sequence length of 512 tokens due to absolute positional embeddings. However, XLNet integrates the relative positional encoding, as proposed in Transformer-XL~\cite{Dai2019}.
Therefore, XLNet's architecture is, in theory, not bound to a maximum sequence length. 
However, a custom pretraining is out of scope for this research, and the publicly available pretrained models of XLNet have the same 512 token limit as BERT.
It remains unknown how the length of the processed sequence affects the classification task.
From \cite{Schwarzer2016}, we know that the performance of similarity measures peaks at 450 words 
since the introduction section in Wikipedia articles presumably contains all essential information. 
Other sections might add only noise and make it harder to encode relevant semantic information from the articles. 
Thus, we evaluate the Transformers using 128, 256, and 512 tokens (Section~\ref{ssec:max-seq-len-results}).

\subsubsection{Concatenation} \label{sssec:concat}

Doc2vec, GloVe, and the Siamese models concatenate the separately encoded document vectors $\vec{d_s}$ and $\vec{d_t}$. %
In the literature, there is no widely accepted concatenation method. For instance, Conneau et al.~\cite{Conneau2017} use $[u;v;|u-v|;u*v]$ for sentence embedding, while Sentence-BERT~\cite{Reimers2019} presents $[u;v;|u-v|]$ as the best method. 
In Section~\ref{ssec:concat-results}, we test the following variations:

\begin{itemize}
\item $[u;v]$ Concatenation of the two vectors $u$ and $v$;
\item $[u;v;|u-v|]$ and absolute value of element-wise difference;
\item $[u;v;|u-v|;u*v]$ and element-wise product.
\end{itemize}

\subsection{Implementation}
\label{ssec:implementation}
All experiments with Doc2vec and AvgGloVe can be run on CPU in less than 15 minutes using the Gensim~\cite{gensim} and Scikit-learn~\cite{scikit-learn} framework.
Before training the Doc2Vec model, Gensim preprocesses\footnote{\url{https://radimrehurek.com/gensim/utils.html\#gensim.utils.simple\_preprocess}} the plain-text from the Wikipedia articles.
For AvgGloVe, the individual words occurring in the article text are extracted with Scikit-learn's CountVectorizer\footnote{\url{https://scikit-learn.org/stable/modules/generated/sklearn.feature\_extraction.text.CountVectorizer.html}} including English stop word removal.
The Transformer models require a GPU as hardware. We rely on HuggingFace's PyTorch implementation~\cite{Wolf2019} of BERT and XLNet. The training time for a single epoch on a GeForce GTX 1080 Ti (11 GB) ranged from less than 10 minutes for vanilla BERT-128 (simplest Transformer architecture), to 55 minutes for Siamese XLNet-512 (most complex Transformer architecture). As suggested in~\cite{Devlin2019}, the Transformer training is performed with batch size $b=4$, dropout probability $d=0.1$, learning rate $\eta=2^{-4}$ (Adam optimizer) and 4 training epochs. If not otherwise stated, the default settings of the frameworks were used.
The evaluation is conducted as stratified k-fold cross-validation with $k=4$ and 24,126 training, and 8,041 test samples (the class distribution remains identical for each fold).
The source code, dataset, and trained models are publicly available on Zenodo\footnote{\url{https://doi.org/10.5281/zenodo.3713183}},  GitHub\footnote{\url{https://github.com/malteos/semantic-document-relations}} and as a demo on Google Colab\footnote{\url{https://ostendorff.org/r/jcdl2020-colab}}.
\section{Results}
\label{sec:results}
Our results are divided in: overall, sequence length, concatenation, relation classes, and manual sample examination. These five subsections move from a high-level perspective to a detailed investigation of the main aspects that most contributed to our findings.
\subsection{Overall}
\label{ssec:overall-results}

The empirical results of the tested systems and hyperparameters are presented in Table~\ref{tab:results}.
Vanilla BERT-512 yields the best micro average F1-Score with 0.933, followed by its 256 length size model, with 0.930. 
The second-best model is the vanilla XLNet-512 with 0.926 F1 and a statistically significant lower score compared to vanilla BERT-512 (95\% confidence interval). 
The vanilla Transformers generally outperform their Siamese counterparts.
Siamese BERT (0.870 F1) and Siamese XLNet (0.870 F1) do not achieve the same performance as their vanilla architectures for the same 128 sequence length size, with scores of 0.920 (BERT-128) and 0.914 (XLNet-128) respectively. 
The shared contextual information during the encoding of document pairs most likely yields the better performance of vanilla Transformers. 
AvgGloVe (0.875 F1) outperforms Siamese BERT and Siamese XLNet, which makes AvgGloVe preferable over Siamese Transformers since AvgGloVe requires only a fraction of the computing resources and runs on commodity hardware. 
With an F1-score of 0.845 at its best configuration, Doc2vec is the worst performing model.
In summary, we consider the results of AvgGloVe and vanilla BERT as most promising for future application scenarios.
We hypothesize that an F1-score of above 0.90 is already suitable enough for LRS.
Especially, expert users would tolerate some misclassifications in favor of otherwise undiscoverable information.
This would be the case for target documents that are considered to be dissimilar to the seed with existing methods but are found to have semantic relation with the help of our methods.

\begin{table}[]
\small
\caption{\label{tab:results} Results as micro avg. F1-score with standard deviation in 4-fold cross-validation for all system configurations including full-text document embeddings from GloVe and Doc2vec, and vanilla and Siamese Transformers (BERT-base and XLNet-base). Vanilla BERT-512 performs best.}
\begin{tabular}{l|lc|rr} %
\textbf{Model} & \textbf{Seq.} & \textbf{Concatenation}  &  F1 &   Std. \\
\midrule
\multirow{3}{*}{AvgGloVe} & \multirow{3}{*}{-} & $u;v$ &  0.863 & $\pm$ 0.0040 \\
              &     & $u;v;|u-v|$ &  0.871 & $\pm$ 0.0045 \\
              &     & $u;v;|u-v|;u*v$ &  0.875 & $\pm$ 0.0036 \\
\cline{1-5}
\cline{2-5}
\multirow{3}{*}{Doc2vec} & \multirow{3}{*}{-} & $u;v$ &  0.838 & $\pm$ 0.0049 \\
              &     & $u;v;|u-v|$ &  0.836 & $\pm$ 0.0048 \\
              &     & $u;v;|u-v|;u*v$ &  0.845 & $\pm$ 0.0019 \\
\cline{1-5}
\cline{2-5}
\multirow{9}{*}{Siamese BERT} & \multirow{3}{*}{128} & $u;v$ &  0.844 & $\pm$ 0.0025 \\
              &     & $u;v;|u-v|$ &  0.859 & $\pm$ 0.0080 \\
              &     & $u;v;|u-v|;u*v$ &  0.856 & $\pm$ 0.0102 \\
\cline{2-5}
              & \multirow{3}{*}{256} & $u;v$ &  0.851 & $\pm$ 0.0046 \\
              &     & $u;v;|u-v|$ &  0.860 & $\pm$ 0.0137 \\
              &     & $u;v;|u-v|;u*v$ &  0.862 & $\pm$ 0.0090 \\
\cline{2-5}
              & \multirow{3}{*}{512} & $u;v$ &  0.846 & $\pm$ 0.0050 \\
              &     & $u;v;|u-v|$ &  0.860 & $\pm$ 0.0087 \\
              &     & $u;v;|u-v|;u*v$ &  0.870 & $\pm$ 0.0067 \\
\cline{1-5}
\cline{2-5}
\multirow{9}{*}{Siamese XLNet} & \multirow{3}{*}{128} & $u;v$ &  0.855 & $\pm$ 0.0075 \\
              &     & $u;v;|u-v|$ &  0.869 & $\pm$ 0.0061 \\
              &     & $u;v;|u-v|;u*v$ &  0.867 & $\pm$ 0.0068 \\
\cline{2-5}
              & \multirow{3}{*}{256} & $u;v$ &  0.856 & $\pm$ 0.0106 \\
              &     & $u;v;|u-v|$ &  0.869 & $\pm$ 0.0071 \\
              &     & $u;v;|u-v|;u*v$ &  0.870 & $\pm$ 0.0078 \\
\cline{2-5}
              & \multirow{3}{*}{512} & $u;v$ &  0.856 & $\pm$ 0.0110 \\
              &     & $u;v;|u-v|$ &  0.860 & $\pm$ 0.0179 \\
              &     & $u;v;|u-v|;u*v$ &  0.864 & $\pm$ 0.0096 \\
\cline{1-5}
\cline{2-5}
\multirow{3}{*}{Vanilla BERT} & 128 & - &  0.920 & $\pm$ 0.0028 \\
              & 256 & - &  0.930 & $\pm$ 0.0042 \\
              & 512 & - &  \textbf{0.933} & $\pm$ 0.0039 \\
\cline{1-5}
\multirow{3}{*}{Vanilla XLNet} & 128 & - &  0.914 & $\pm$ 0.0065 \\
              & 256 & - &  0.914 & $\pm$ 0.0023 \\
              & 512 & - &  0.926 & $\pm$ 0.0016 \\
\end{tabular}

\end{table}

\subsection{Sequence Length}
\label{ssec:max-seq-len-results}

\begin{figure}[h]
\centering
\includegraphics[clip,width=0.45\textwidth]{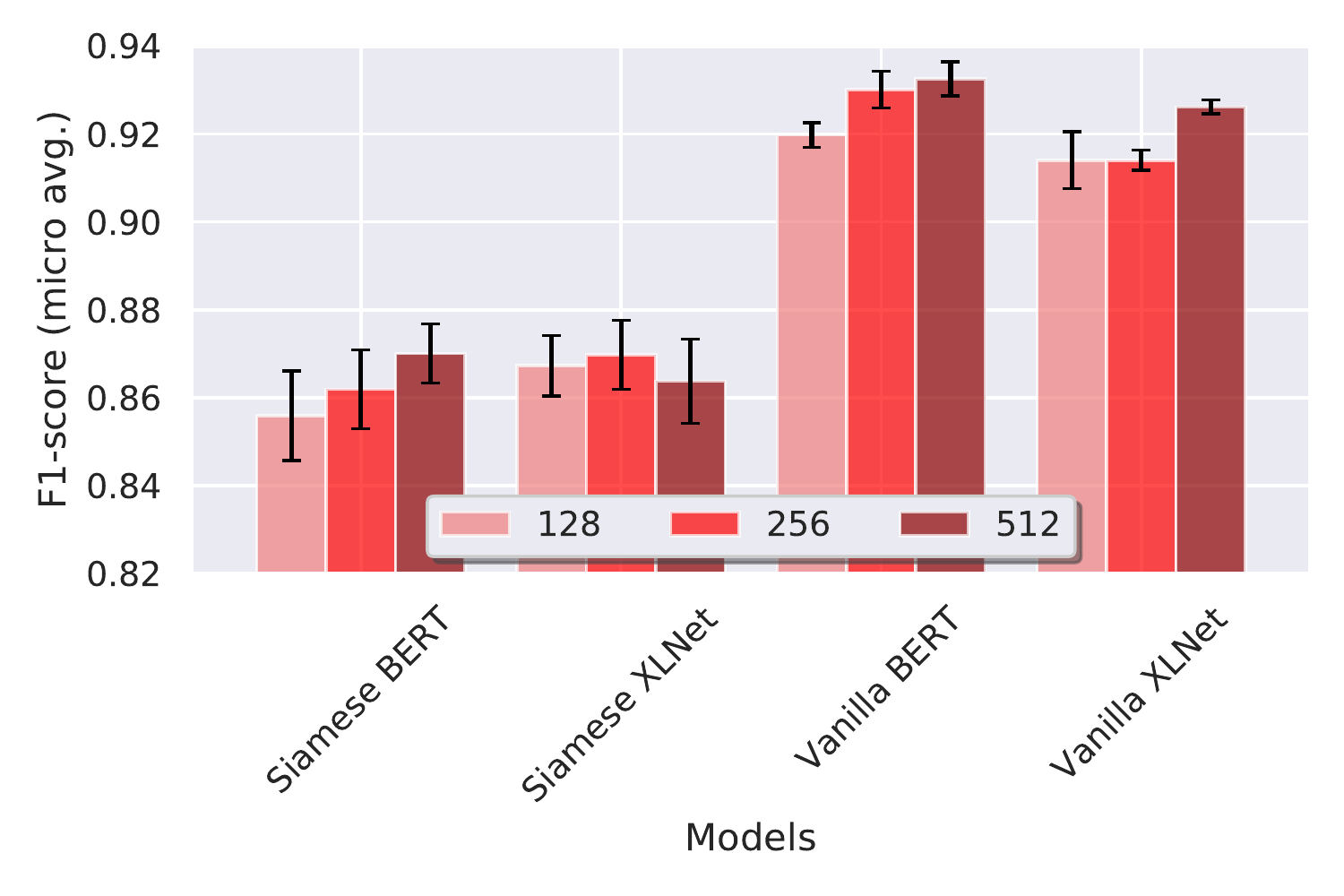}
\caption{\label{fig:max-seq-len}Performance of vanilla and Siamese Transformers w.r.t. sequence length. Siamese models use $[u;v;|u-v|;u*v]$ as concatenation. Aside from Siamese XLNet, 512 tokens achieve the best F1-scores for all models.}
\end{figure}

As explained in Section~\ref{ssec:hyperparameters}, we are particularly interested in the effect of the sequence length on the Transformer models. 
To illustrate this effect, Figure~\ref{fig:max-seq-len} shows the comparison of Siamese BERT, Siamese XLNet, vanilla BERT, and vanilla XLNet with respect to their sequence length (i.e., 128, 256, and 512). 
In this comparison, the Siamese models use the best performing concatenation method, which is $[u;v;|u-v|;u*v]$. 
Our findings reveal that longer sequences are related to better results. 
For all models, except Siamese XLNet, the highest F1-score is achieved with 512 tokens and the second-highest with 256 tokens. 
One could think this outcome is to be expected.
However, in \cite{Schwarzer2016}, the performance of text- and link-based document similarity measures declines for Wikipedia articles with more than 450 words. 
When comparing Siamese with vanilla Transformers, the vanilla models work with only half of the sequence length to encode one document of the pair. 
In vanilla Transformers, the document pairs share the sequence length, while in Siamese Transformers each document has its own Transformer sub-network (sequence length).
For example, a vanilla 128-Transformer would use only 62 or 63 tokens of each document (three tokens are reserved for special tokens as Figure~\ref{fig:transformer} shows). 
Thus, the small performance difference within vanilla BERT with 512 tokens (0.933 F1), 256 tokens (0.930 F1), and 128 tokens (0.920 F1) is remarkable. %
Moreover, the performance differences should be considered relative to the higher computation expenses of longer sequences. 

\subsection{Concatenation} \label{ssec:concat-results}

\begin{figure}[h]
\centering
\includegraphics[clip,width=0.45\textwidth]{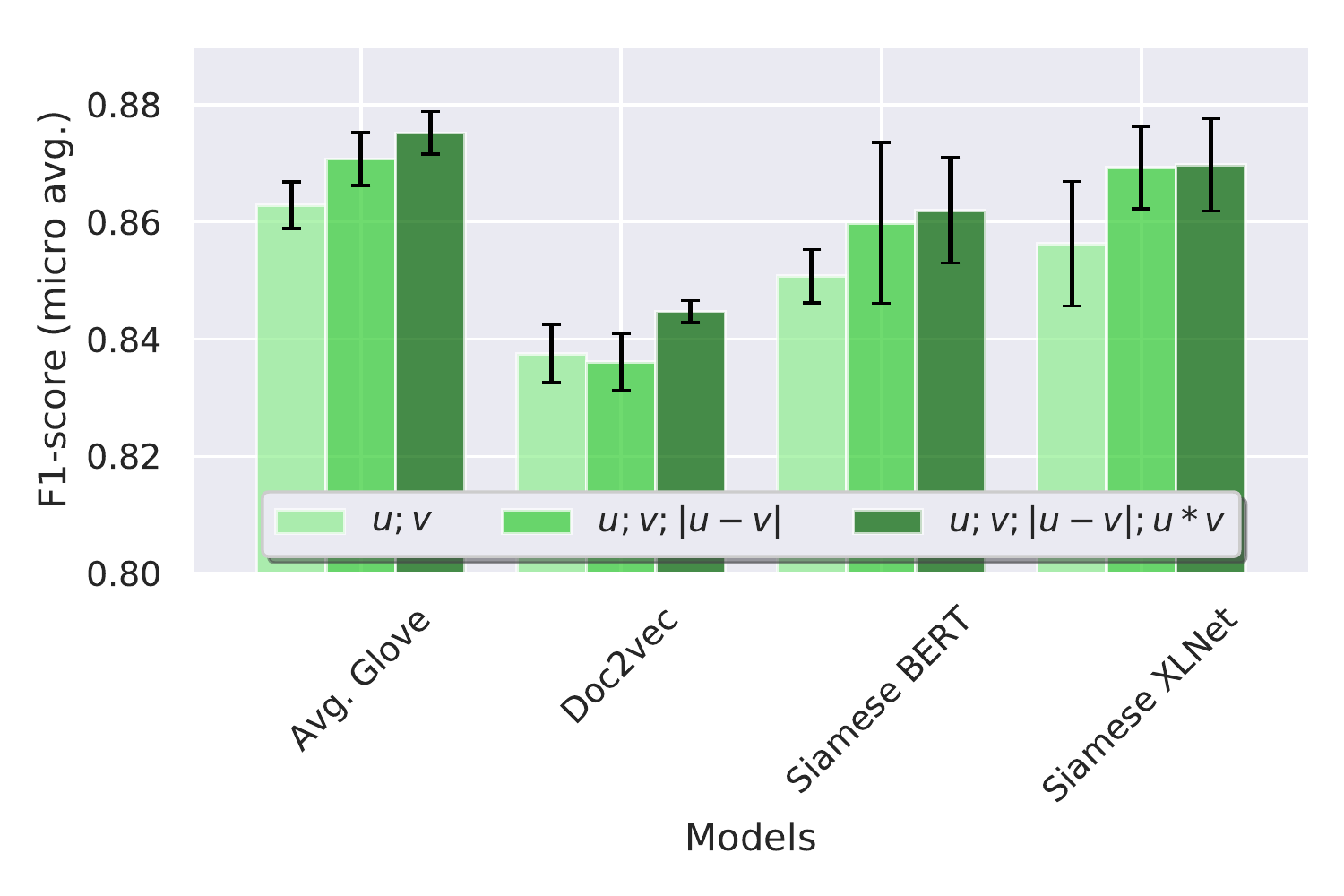}
\caption{\label{fig:concat}Results of the full-text document embeddings and Siamese Transformer-512 models w.r.t. concatenation.}
\end{figure}

Aside from the sequence length, we also analyzed the different concatenation methods in AvgGloVe, Doc2vec, and the Siamese models (Figure~\ref{fig:concat}).
All models achieve the highest F1-score when the concatenation with an element-wise difference and product is used ($[u;v;|u-v|;u*v]$). 
Furthermore, we confirmed the results of Reimers and Gurevych~\cite{Reimers2019}, i.e., the most crucial component is the element-wise difference $|u-v|$. 
Only for Doc2vec the element-wise difference decreases the performance in comparison to the simple concatenation. However, this performance decrease is marginal and within standard deviation. In general, the element-wise difference measures the distance between the dimensions of the two document vectors and, thus, ensures that similar pairs are closer to each other than dissimilar pairs. This effect is evident for Siamese BERT and Siamese XLNet, for which the element-wise difference yields the most substantial performance improvement. 
On the contrary, the element-wise product adds only a small improvement to our models.

\subsection{Relation Classes} \label{ssec:classes-results}

\begin{table*}[h]
\small
\caption{\label{tab:results-by-relation} Results for precision (P), recall (R), F1-score, and sample count in test data w.r.t. relation classes. Evaluated systems are AvgGloVe, Siamese BERT, vanilla BERT and vanilla XLNet. The results of other models are published along with the code. }

\begin{tabular}{l|rrr|rrr|rrr|rrr|r}

\textbf{Model} & \multicolumn{3}{c}{\textbf{AvgGloVe}} & \multicolumn{3}{c}{\textbf{Siamese BERT-512}} & \multicolumn{3}{c}{V\textbf{anilla BERT-512}} & \multicolumn{3}{c}{\textbf{Vanilla XLNet-512}} & {} \\ \hline
Relation class &  P & R & F1 &    P & R & F1 &    P & R & F1 &     P & R & F1 & Samples\\

\hline

country of citizenship 
&      0.963 &  0.983 &    0.973 
&      0.956 &  \textbf{0.996} &    0.976 
&      \textbf{0.993} &  \textbf{0.996} &    \textbf{0.994}
&      0.989 &  \textbf{0.996} &    0.993 
&     909 \\

\rowcolor{Gray}
different from         
&        0.856 &  0.843 &    0.849 
&        0.872 &  0.899 &    0.885 
&         \textbf{0.971} &   0.931 &     \textbf{0.950}
&        0.969 &   \textbf{0.933} &     \textbf{0.950}
&    1012 \\

educated at            
&      0.683 &  0.729 &    0.703 
&        0.730 &  0.740 &    0.734 
&        0.759 &  \textbf{0.900} &    \textbf{0.817}
&         \textbf{0.774} &  0.759 &    0.763 
&     450 \\

\rowcolor{Gray}
employer               
&      0.662 &  0.620 &    0.639 
&        0.639 &  \textbf{0.769} &    0.695 
&        \textbf{0.892} &  0.653 &    \textbf{0.740} 
&         0.711 &  0.748 &    0.725 
&     389 \\

facet of               
&      0.786 &  0.781 &    0.782 
&        0.839 &  0.785 &    0.810 
&        \textbf{0.916} &  \textbf{0.908} &    \textbf{0.911} 
&         0.888 &  0.904 &    0.896 
&     336 \\

\rowcolor{Gray}
has effect             
&      0.644 &  0.606 &    0.620 
&        0.626 &  0.468 &    0.502 
&        \textbf{0.783} &  0.614 &    0.683 
&         0.768 &  \textbf{0.658} &    \textbf{0.704}
&     175 \\
has quality            
&      0.694 &  0.682 &    0.687 
&        0.662 &  0.619 &    0.639 
&        0.718 &  0.797 &    0.749 
&         \textbf{0.763} &  \textbf{0.799} &    \textbf{0.774} 
&     256 \\

\rowcolor{Gray}
opposite of            
&      0.672 &  0.666 &    0.667 
&        0.540 &  0.791 &    0.640 
&        0.761 &  0.763 &    0.756 
&         \textbf{0.773} &  \textbf{0.835} &    \textbf{0.795} 
&     232 \\

symptoms               
&      \textbf{0.887} &  0.932 &    0.908 
&        0.827 &  0.969 &    0.892 
&        0.872 &  0.973 &    \textbf{0.920} 
&         0.864 &  \textbf{0.984} &    0.919 
&     263 \\

\rowcolor{Gray}
\textit{none}                   
&      0.943 &  0.940 &    0.942 
&        0.955 &  0.897 &    0.925 
&        0.978 &  \textbf{0.981} &    \textbf{0.979} 
&         \textbf{0.979} &  0.968 &    0.973 
&    4021 \\ \hline
micro avg               
&      0.875 &  0.875 &    0.875 
&        0.870 &  0.870 &    0.870 
&        \textbf{0.933} &  \textbf{0.933} &    \textbf{0.933} 
&         0.926 &  0.926 &    0.926 
&       8043 \\

macro avg              
&      0.779 &  0.778 &    0.777 
&        0.764 &  0.793 &    0.770 
&        \textbf{0.864} &  0.852 &    \textbf{0.850} 
&         0.848 &  \textbf{0.858} &    0.849 
&    8043 \\

\end{tabular}

\end{table*}

We selected \relCount diverse Wikidata properties to explore how the systems would respond to the individual challenges of each property. %
Table~\ref{tab:results-by-relation} presents precision, recall, and F1-score of the best four systems for the different model categories. 
Each score is the mean over the 4-fold cross-validation (cf. Table~\ref{tab:results} for standard deviation). 
AvgGloVe and Siamese BERT use  $[u;v;|u-v;u*v]$ as concatenation method, and all Transformer models (Siamese BERT, vanilla BERT, and vanilla XLNet) use the 512 sequence length. 
The best relation classes in terms of performance are \textit{country of citizenship}, \textit{none} (negative samples), and \textit{different from}, whereas the classes \textit{employer}, \textit{has quality}, \textit{has effect} yield the lowest scores. 
Given that the best performing classes are also over-represented in terms of sample count, the outcome suggests that other classes only need more training data. 
Still, the comparison of the \textit{employer} class (389 test samples, vanilla BERT 0.740 F1) and \textit{facet of} (336 test samples, vanilla BERT 0.911) reveals that the performance difference is also due to the diverse requirements of classes themselves.

\begin{figure}[h]
\centering
\includegraphics[clip,width=0.49\textwidth]{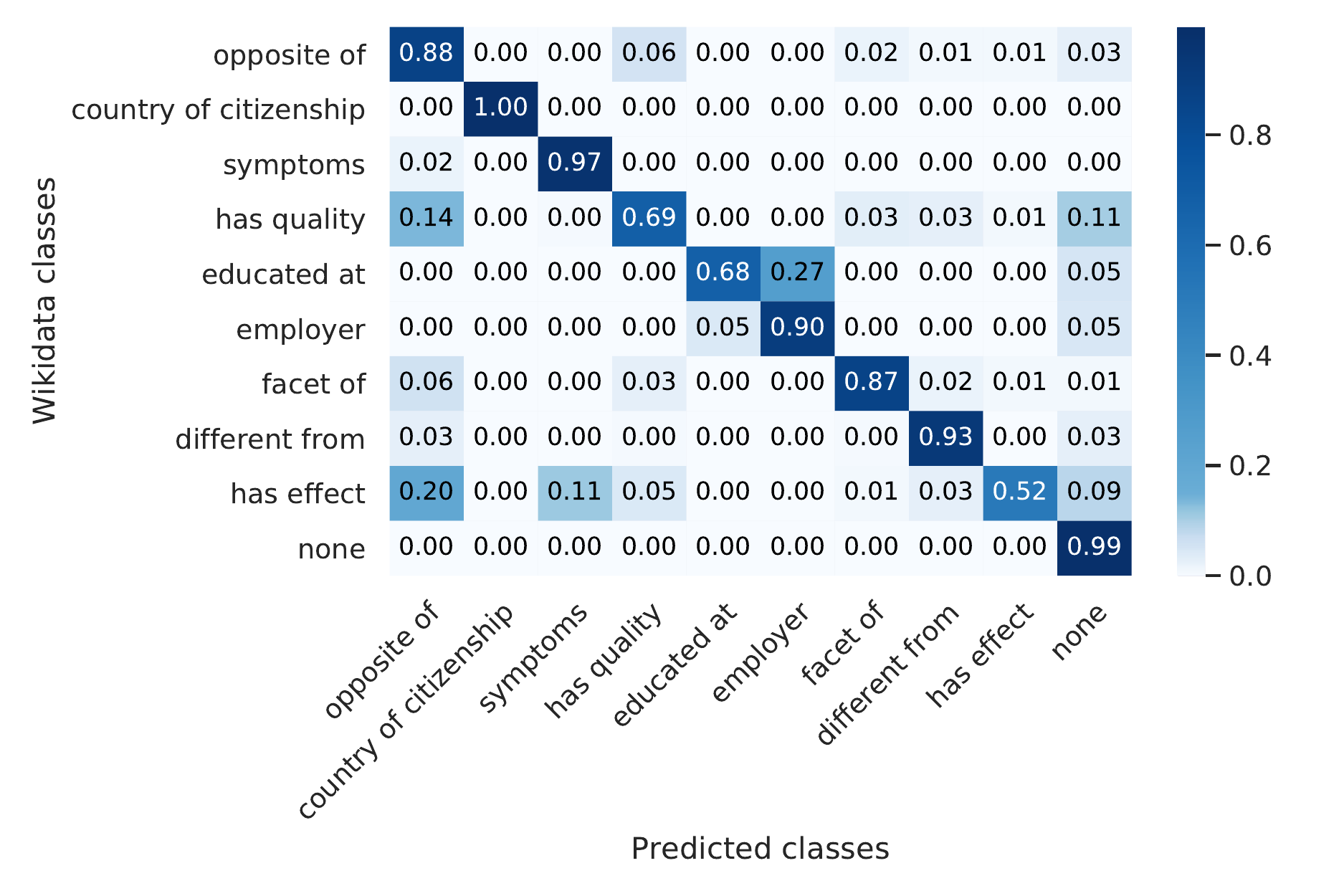}
\caption{\label{fig:confusion}Confusion matrix for the predicted and Wikidata classes of vanilla BERT. The relation count is normalized. The most frequent confusion is found with the \textit{educated at} and \textit{employer} class for 27\% of the test samples. }
\end{figure}

The superiority of vanilla BERT is also present in the class-specific evaluation scenario, although it is outperformed by vanilla XLNet for three relation classes with a small number of samples (\textit{has effect}, \textit{has quality} and \textit{opposite of}). 
In AvgGloVe, \textit{symptoms} has the highest precision score, which is probably caused by AvgGloVe being able to utilize the full-text of articles in contrast to the Transformer models. Medical articles, like \textit{Alcoholism} (Example 9 in Table~\ref{tab:examples}), contain a section ``Signs and symptoms'' in which their symptoms are listed. 
However, such a section is not part of the 512 Transformer tokens.
When comparing precision and recall for all classes, both scores are mostly balanced. There is only one striking exception for vanilla BERT. 
For \textit{employer}, the precision score of 0.829 is higher than the recall of 0.653, while for \textit{educated at} the opposite occurs, with a precision of 0.759 and recall of 0.900, but in a smaller magnitude. 
A reason for this outcome is that \textit{employer} is often confused with \textit{educated at} as Figure~\ref{fig:confusion} shows.

The confusion matrix in Figure~\ref{fig:confusion} depicts which classes are most often confused with each other. 
The predicted classes are taken from the vanilla BERT-512 system, whereby the number of true and predicted classifications is normalized to make the different classes comparable. 
With 27\% of the test sample, \textit{educated at} and \textit{employer} are the most mistaken relation classes in our experiments. 
We see this outcome because both relation classes connect persons and organizations, and we assume it is harder for the classifier to tell the relations apart.
For instance, \textit{Albert Einstein} could be employed or educated at \textit{ETH Zurich} (Figure~\ref{fig:wikidata-usecase}).
The misclassification between different relations is also found in \textit{opposite of}, \textit{has quality} and \textit{has effect}, which we conclude is because of similar reasons. In particular, the \textit{opposite of} relation connects various types of articles.
\subsection{Manual Sample Examination}
\label{ssec:examples}
To validate our empirical findings, we manually examine the prediction from vanilla BERT-512 with a focus on errors (Table~\ref{tab:examples}).
Examples 1 and 2 show a desired classifier despite one misclassification according to Wikidata. 
While \textit{Armenia} is correctly identified as \textit{Rudolf Muradyan's} \textit{country of citizenship}, \textit{Brazil} is not recognized. 
However, \textit{Brazil} is also not mentioned in \textit{Rudolf Muradyan's} Wikipedia article. 
The Wikidata statement is not reflected in the Wikipedia article, which states \textit{Muradyan} as \textit{Armenian} only. 
Consequently, both predictions would be correct when only considering the article text. %
Two errors are exemplified in 3 and 4. 
Even though \textit{Zaki Naguib Mahmoud's} article explicitly expresses the \textit{educated at} relation with the sentence ``Mahmoud was educated at Cairo University'', \textit{Cairo University} is classified as his \textit{employer}. 
Despite not being mentioned in \textit{Mahmoud's} article, \textit{King's College London} is also wrongly classified as his \textit{employer}. 
In example 5, \textit{Light} is incorrectly classified as the quality of \textit{Darkness}, not as opposite of it. 
Still, \textit{opposite of} is the class with the second-highest probability. 
Example 6 shows the \textit{Mexican Revolution} as \textit{different from} the \textit{Mexican War of Independence}, which would be clear to a human user since the Wikipedia article contains banner ``Not to be confused with the Mexican War of Independence.''. 
However, this banner is missing in the Wikipedia dump and, thus, is not available to the classifier. 
Many shared terms and vocabulary make their classification hard to predict for the \textit{different from} relation.
Examples 7-9 are not discussed, but similarly illustrate the classifier's performance.

Our manual examination confirms the overall results. 
Most relations are correctly identified, %
while some relations are missing even if they are explicitly mentioned in the text.
An analysis of the inner Transformer components~\cite{Clark2019} is a subject for future work.

\begin{table*}[h]
\small
\caption{\label{tab:examples} Example relations between Wikipedia article pairs (seed and target) as defined by Wikidata and as predicted by Vanilla BERT-512 with the first and second highest probability. Correct predictions are marked with~\checkmark.
}
\begin{tabular}{c|ll|lll}

\textbf{ID} & \textbf{Seed} & \textbf{Target}  & \textbf{Wikidata Relation}                  & \textbf{1st Prediction}            & \textbf{2nd Prediction}        \\ \hline
1                            & Rudolf Muradyan                   & Brazil                           & country of citizenship & none                  & country of citizenship~\checkmark   \\

\rowcolor{Gray}
2                            & Rudolf Muradyan                   & Armenia                          & country of citizenship & country of citizenship~\checkmark  & none                    \\
3                            & Zaki Naguib Mahmoud               & Cairo University                 & educated at           & employer              & educated at~\checkmark              \\

\rowcolor{Gray}
4                            & Zaki Naguib Mahmoud               & King's College London            & educated at           & employer              & educated at~\checkmark              \\

5                            & Light                             & Darkness                         & opposite of           & has quality           & opposite of~\checkmark              \\

\rowcolor{Gray}
6                            & Mexican Revolution                & Mexican War of Independence      & different from        & has effect            & symptoms                \\

7                            & History of blogging               & Blog                             & facet of              & opposite of           & different from          \\

\rowcolor{Gray}
8                            & Iced tea                          & Ice-T                            & different from        & none                  & different from~\checkmark           \\
9                            & Alcoholism                        & Cirrhosis                        & has effect            & has effect~\checkmark             & none                   
\end{tabular}

\end{table*}

\section{Discussion} \label{sec:discussion}

Given the results in Table~\ref{tab:results}, we can state that vanilla Transformers outperform all other methods. 
Rather unexpected is that BERT generally achieves slightly better results than XLNet. 
According to Yang et al.~\citep{Yang2019}, XLNet surpasses BERT on the related GLUE benchmark~\cite{Wang2019}, so we were expecting a similar outcome. %
We hypothesize that this difference may be attributed to two reasons, pretraining on different corpora, and smaller models compared to \citep{Yang2019}. 
We use the \texttt{BASE}, not the \texttt{LARGE} versions of the pretrained models used by Yang et al.~\citep{Yang2019}. 
Furthermore, the published XLNet \texttt{BASE} model we considered is pretrained on different data than the one in Yang et al.~\citep{Yang2019}\footnote{See \url{https://github.com/zihangdai/xlnet\#released-models} ``\textit{This model (XLNet-Base) is trained on full data (different from the one in the paper)}''.}.
In contrast to BERT, XLNet is pretrained on Web corpora in addition to Wikipedia and the BooksCorpus~\cite{Zhu2015}.
The almost exclusive pretraining on Wikipedia most likely causes BERT to surpass XLNet. 
The effect of domain-specific pretraining on the performance of the language model has already been shown~\cite{Beltagy2019}. 

Our evaluation also shows that the Siamese networks cannot capture the semantic relations as good as vanilla Transformers. 
In Siamese models, the encoding of the seed document does not affect the target, and vice-versa. 
Only the MLP is exposed to the documents as a pair in the form of the concatenated document vectors. 
During the encoding phase, the relation between the documents plays no role. 
On the contrary, the Multi-Head-Attention mechanism in the vanilla Transformers allows attending on the two documents simultaneously. 
As the results suggest, this ability is crucial for the pairwise document classification. 
The Siamese models are also outperformed by the computationally less expensive AvgGloVe. 
At a general level, the Siamese models are very similar to AvgGloVe (and Doc2vec), since they derive two document vectors and classify their concatenation. 
So the performance of the method ultimately depends on its ability to encode the documents. 
Arora et al.~\cite{Arora2017} have shown that the weighted average of word vectors can outperform more sophisticated methods. 
AvgGloVe benefits from the fact that it utilizes the full-text article in contrast to the Transformers, which use only the 512 first tokens of the article text.
As a result, AvgGloVe is a reasonable method for real production scenarios, in which computational resources are critical concerns.
In production scenarios, one would also avoid classifying all possible $n^2$ document pairs.
Instead, evidently unrelated pairs must be filtered out with traditional similarity measures at first.

Regarding the different relation classes, almost all results present reasonable performance. Moreover, complex relation classes like  \textit{facet of} or \textit{has effect}, yield promising results, since they are attractive for the recommender system use case. As the example 1 and 2 shows in Table~\ref{tab:examples}, current systems already reveal wrong or contradicting information between Wikidata and Wikipedia. The results suggest that increasing the sequence length beyond the 512 tokens could further improve the Transformer models. Higher sequence length is already possible with XLNet's architecture, but it would require a pretraining step with longer sequences.

\paragraph{From Relations to Recommendations} \label{sssec:rel2rec}
Classifying the document relations is not a purpose on its own. We envision recommender systems as an example of a downstream task. 
The obtained relations can be used for diverse or focused recommendations. 
As the relations describe different facets of the seed document, one could diversify the recommendations. 
Choosing the recommendations from documents connected with different relation classes to the seed document would ensure diversity. 
In Figure~\ref{fig:wikidata-usecase}, the \textit{German Empire} and \textit{ETH Zurich} can be considered as diverse recommendations, since they present different aspects of \textit{Albert Einstein}, i.e., his citizenship and education. 
When considering documents that are connected to a seed (i.e., one common document) over two edges (i.e., different relations), recommendations focusing on specific aspects are more feasible. 
Diverse and focused recommendations could be especially suitable for scenarios in which different perspectives are required for the same seed. 
In contrast to user-based recommender systems, content-based approaches usually struggle to account for specific preferences from their users. 
One way to respect different information requirements would be to suggest alternative recommendation sets that are focused on specific aspects. 
In the example of \textit{Albert Einstein}, shown in Figure~\ref{fig:wikidata-usecase}, focused recommendation sets could include articles about people with the same citizenship or the same educational backgrounds.
The intersection of relations would even allow finding people with the same citizenship but different educational background.
The classification of the document relations, as demonstrated in our experiments, is the foundation for such recommendations.

\paragraph{Generalization}

Given the long-term goal of applying the tested methods on non-encyclopedic corpora, the question arises whether our findings are generalizable.
We acknowledge that Wikipedia is a presumable a simpler corpus compared to other literature domains like research papers.
Wikipedia articles represent distinct entities and most relations are explicitly expressed in the article text.
However, even research papers express semantic relations in their abstracts, e.g., ``we used X'' or ``we found Y''.
Accordingly, we hypothesize that our systems would yield worse but still satisfactory results under comparable conditions (size of training data, pretraining on in-domain corpus etc.).
A reference value would be the F1-score of 0.65, which was achieved by SciBERT on the related task of citation intent classification~\cite{Beltagy2019}.
While the effort for the unsupervised pretraining of a language model is reasonable, we recognize the annotation of sufficient training data for other corpora is one of the most challenging tasks.
After all, even annotations can be solved efficiently as Chan et al.'s crowdsourcing approach demonstrates~\cite{Chan2018}.
We are confident that our results are transferable to other domains.

\section{Conclusion and Future Work} \label{sec:conclusion}

This paper introduces the pairwise document classification to determine semantic relations between documents as an underlying task to advance LRS and other information retrieval applications. 
We elaborate on why document similarity measures do not account for the heterogeneous semantics of extensive documents and argue that similarity needs a context which defines to what it relates. 

The task of finding semantic document relations is implemented as a multi-class classification of document pairs. 
We demonstrate the viability of this approach with a new proposed dataset of 32,168 Wikipedia article pairs and Wikidata properties that define semantic relations among these articles. 
In an empirical study, we implement six different models AvgGloVe, Doc2vec, Siamese BERT, Siamese XLNet, vanilla BERT and vanilla XLNet, and evaluate them under different settings regarding the concatenation method and sequence length (Table~\ref{tab:results}). 
Our evaluation indicates a sequence length of 512 tokens as the best performing sequence limit for the Siamese and vanilla Transformer models. 
In addition, we identify $[u;v;|u-v|;u*v]$ as the best concatenation method for AvgGloVe, Doc2vec and the Siamese Transformer models. 
With the manual sample examination and our evaluation for different relation classes, we show the behavior of the classifiers when exposed to different input data and provide an analysis over different perspectives.
Moreover, the manual analysis confirms our empirical results.

Our findings suggest that pairwise document classification is a solvable task using existing techniques. 
Even abstract semantic relations, like \textit{facet of}, yield a considerable high F1-score. 
This outcome motivates us to investigate the semantic relations between documents of other literature domains, primarily scientific papers.
We envision a system that enables users to explore scientific literature in an analogical manner. 
For instance, users could retrieve other research papers with a similar methodology but different result.
Analogies could be even found with programmatic and SPARQL-like queries.
To develop such a system, the Open Research Knowledge Graph~\cite{Jaradeh2019} could be utilized as the scientific equivalence of Wikidata, while research paper from any open digital library, e.g., arXiv, would correspond to Wikipedia articles. 
Lastly, the presented Wikipedia and Wikidata dataset also facilitate the evaluation of methods in terms of required training data. The estimation the necessary amount of data would a prerequisite for looking into other domains.

\begin{acks}

The research presented in this article is funded by the German Federal Ministry of Education and Research (BMBF) through the project QURATOR (Unternehmen Region, Wachstumskern, no.~03WKDA1A). 

\end{acks}

\bibliographystyle{ACM-Reference-Format}
\bibliography{references}

\end{document}